\documentclass[12pt,a4paper]{article}

\usepackage{fullpage}
\usepackage{amsfonts}
\usepackage{amssymb}
\usepackage{lettrine}
\usepackage{graphicx}
\usepackage{amsmath}
\usepackage{amsthm}
\usepackage{url,doi}
\usepackage{hyperref}

\newtheorem{definition}{Definition}
\newtheorem{theorem}[definition]{Theorem}

\newtheorem{corollary}[definition]{Corollary}

\newcommand{\ket}[1]{{\left\lvert{#1}\right\rangle}}

\newcommand{\Tr}{{{\mathop{\rm Tr}}}}
\newcommand{\HPP}{\mbox{\rmfamily\textsc{HPP}}}
\newcommand{\HPFGP}{\mbox{\rmfamily\textsc{HPGP}}}

\def\Z{{\mathbb Z}}

\def\F{{\mathbb F}}

\title{Polynomial time quantum algorithms for certain bivariate hidden polynomial problems}

\author{
Thomas Decker\thanks{Centre for Quantum Technologies, 
National University of Singapore, Singapore 117543
({\tt t.d3ck3r@gmail.com}).} 
\and
Peter H{\o}yer\thanks{
Department of Computer Science and 
Institute for Quantum Science and Technology,
University of Calgary, 2500 University Drive N.W.,
Calgary, Alberta, Canada, T2N 1N4
({\tt hoyer@ucalgary.ca}).}
\and
G{\'a}bor Ivanyos\thanks{Institute for Computer Science and Control, Hungarian Academy of Sciences, 
Budapest, Hungary 
({\tt Gabor.Ivanyos@sztaki.mta.hu}).} 
\and
Miklos Santha\thanks{LIAFA, Univ. Paris 7, CNRS, 75205 Paris, France;  and 
Centre for Quantum Technologies, National University of Singapore, 
Singapore 117543 ({\tt miklos.santha@liafa.jussieu.fr}).}
}

\begin{document}
\maketitle

\begin{abstract}
We present a new method for solving the hidden polynomial graph
problem (HPGP) which is a special case of the hidden polynomial problem
(HPP). The new approach yields an efficient quantum algorithm 
for the bivariate HPGP even when
the input consists of several level set superpositions, a more difficult version of the problem
than the one where the input is given by an oracle. For constant degree, the algorithm is polylogarithmic in the size
of the base field.
We also apply the results to give an efficient quantum algorithm for the oracle version of the HPP for
an interesting family of bivariate hidden functions. This family
includes diagonal quadratic forms and elliptic curves. 
\end{abstract}

\section{Introduction}\label{sec:intro}
In the {\em hidden polynomial problem} (\HPP{}) we are given an oracle for a
function of the form ${\cal E}(F(x_1,\ldots,x_n))$, where $F$ is an
unknown polynomial in $n$ variables of degree at most $D$ over the
finite field $\F$ and where ${\cal E}$ is an unknown unique encoding
of elements of $\F$ by binary strings. 
This means that the level sets of the oracle coincide with the level sets of the polynomial. The task is to determine the
polynomial $F$. Obviously, $F$ can only be determined up to a constant
additive term and up to another constant multiplicative
factor. Therefore, we consider polynomials with fixed constant term
(usually zero) and in which another monomial is fixed (usually it has
coefficient 1). In the quantum setting, the oracle is actually a
unitary transformation which maps states of the form
\[
\ket{x_1}\ket{x_2}\cdots\ket{x_n}\ket{0}\quad {\rm to} \quad
\ket{x_1}\ket{x_2}\cdots\ket{x_n}\ket{{\cal E}(F(x_1,\ldots,x_n))}
\,.
\]
We measure the {\em complexity} in terms of the number of bits that are
necessary to describe the polynomial $F$, which is $\Omega(\log |\F|)$ if
$n$ and $D$ are constant.
We say an algorithm is efficient if its
time complexity is polynomial in $\log |\F|$.  We assume that each oracle query can 
be conducted in one time step, where needed.


The \HPP{} was introduced by Childs, Schulman and
Vazirani~\cite{CSV} in an attempt to generalize the study of properties of algebraic sets hidden by black-box functions from linear structures, instantiated by the well-known hidden subgroup problem (HSP), to higher degree cases.
They showed that when the degree of the hidden
polynomial as well as the number of variables is constant, a typical
polynomial can be determined by a polynomial number of
queries. 
 Decker, Ivanyos, Santha and Wocjan~\cite{DISW11}
designed an efficient quantum algorithm for the HPP for multivariate quadratic polynomials over fields of constant characteristic.
In~\cite{DDW09}, Decker, Draisma and Wocjan considered the {\em hidden polynomial graph problem} (\HPFGP{}),
a special case of the \HPP{} where $F(x_1,\ldots,x_n)$ is of the
form $x_n-f(x_1,\ldots,x_{n-1})$ for some polynomial
$f(x_1,\ldots,x_{n-1})$. They showed how to reduce the \HPFGP{}, when
$n$ and the degree are constant, to the bivariate case, that is to
the case of a hidden polynomial $F(x,y)$ of the form $y-f(x)$. They
also gave an efficient quantum algorithm for the bivariate case when
the degree of $f$ is a constant and smaller than the
characteristic of the base field $\F$. The algorithm of~\cite{DDW09}
used a technique analogous to the pretty good measurement framework of
\cite{BCvD05} for solving the HSP in certain
semidirect product groups. 

An explanation for why these two problems can be solved with very 
similar tools was given in~\cite{DISW11} where it was proven that the bivariate \HPFGP{} can be efficiently 
reduced to a special instance of the HSP.
In fact over prime fields this coincides with the problem 
considered in \cite{BCvD05}. Interestingly, there is (almost)
a reduction in the other direction as well: 
based on \cite{BCvD05}, it is shown in
\cite{DISW11} that the hidden subgroup problem in $\Z_p^m\rtimes \Z_p$
can be efficiently reduced to a multidimensional version
of the \HPFGP{}.

In this paper we propose a novel approach for solving a slightly more
difficult version of the bivariate \HPFGP{} in which, rather than an
oracle, we are given quantum states as input. To be more specific, the
input consists of several level set superpositions of the function
$F(u,x)=u-f(x)$, that is, quantum states\footnote{In order to simplify
 notation, we omit the normalization factors of state vectors in
 Section~\ref{sec:intro} and~\ref{sec:linearize}.} of the form
\begin{equation}
\label{eq:levelsup}
\sum_{x \in \F} \ket{w+f(x)}\ket{x}
\,,
\end{equation}
where each state comes with an unknown and possibly different element
$w \in \F$. In the following, we do not assume anything on the
various $w$ corresponding to different input states. This definition
of the \HPFGP{} is more general than the oracle version, because from
the oracle we can easily obtain the level set superpositions of
Eq.~(\ref{eq:levelsup}) for random $w$ according to a distribution
reflecting the frequency of $w$ appearing as a value of $F$. Our main
result is the following.

\begin{theorem}
\label{thm:hpgp}
Let $D$ be a constant and let $f(x)=\sum_{s=1}^D\nu_{s}x^s \in \F[x]$. Then
there is a quantum algorithm which, given $O(1)$ states of the form~(\ref{eq:levelsup})
for various and unknown $w\in \F$, determines the hidden coefficients
$\nu_{1}, \ldots, \nu_D$ efficiently.
\end{theorem}

Actually, the special case of Theorem~\ref{thm:hpgp} where the
characteristic of $\F$ is greater than $D$ could also be proved using
the method of \cite{DDW09}, because it is also a quantum algorithm
that works on states of the form 
(\ref{eq:levelsup}). Our result is nonetheless new for small characteristics.
Observe that 
the HSP is only discussed in the oracle
setting and cannot directly be applied to the states 
(\ref{eq:levelsup}).

As an application, we present an efficient quantum algorithm which
solves 
the oracle version of 
the \HPP{}
for the family of bivariate hidden functions of the form
$F(x,y)=g(y)-f(x)$ where $g(y)$ is a fixed and known non-constant
polynomial of degree $D'$ in $y$ and $f(x)$ is an unknown polynomial
of degree at most $D$ in $x$ with fixed constant term. This class
includes polynomials of the form $y^2-\nu x^2$ (diagonal quadratic
forms) as well as those of the form $y^2-f(x)$ where $f(x)$ is of
degree 3 or higher (elliptic and hyperelliptic curves). 
In contrast to Theorem~\ref{thm:hpgp}, where
the error can be made arbitrarily small, the algorithm for this
theorem has an ingredient  which gives the correct
result with a probability that is bounded by a (small) constant. Hence, we
need the oracle to test the correctness of results.
We show the
following.

\begin{theorem}
\label{thm:hpps}
Let $D$ and $D'$ be constants and let $g(y)=\sum_{s=0}^{D'}\mu_sy^s$
be a fixed polynomial. Furthermore, let $f(x)=\sum_{s=1}^D\nu_{s}x^s$
be a polynomial of degree $D$ with unknown coefficients $\nu_s$.
Then, given a quantum oracle
that maps states of the form
$\ket{y}\ket{x}\ket{0}$ to $\ket{y}\ket{x}\ket{{\cal E}(g(y)-f(x))}$,
the unknown coefficients $\nu_1,\ldots,\nu_D$ can be determined
in polynomial time.
\end{theorem}

The proof of
Theorem~\ref{thm:hpgp} and our method for solving the \HPFGP{} are presented in Section~\ref{sec:linearize}.
The proof of Theorem~\ref{thm:hpps} is given in
Section~\ref{sec:hpps}.

\section{The phase linearization approach}\label{sec:linearize} 
The high level description of our algorithm, that we call {\em phase linearization} is actually quite simple.
The QFT applied to a level set superposition results in a superposition where 
the phases have not only linear but also higher degree exponents.
Our main goal is to eliminate these non-linear exponents, once it is done the inverse QFT yields a linear equation
in the unknown coefficients. To achieve this we will combine several copies of the level set superposition.
The acquired freedom in the composed phase can be used, with the help of an additional register in uniform superposition, 
to make the exponents  linear.

The elimination of the higher degree exponents will be done recursively. For this 
it will be convenient to consider a technical generalization of the \HPFGP{} 
that is suitable for recursion. However, 
before formulating that, we demonstrate phase linearization
in the quadratic case and we also briefly outline an
extension to the cubic case. 

\subsection{The quadratic and cubic cases}
\label{ss-smalldeg}
In this subsection, we assume that $|\F|$ is odd and that our input
consists of several states of the form (\ref{eq:levelsup}) where
$f(x)=\nu x+\mu x^2$, that is we have states
\begin{equation}
\label{eq:levelsup2}
\sum_{x \in \F}\ket{w+\nu x + \mu x^2} \ket{x}
\end{equation}
with various unknown $w\in\F$. The task is to determine the
coefficients $\nu$ and $\mu$. We apply to the first register the quantum Fourier transform
of the field $\F$. This is the unitary transform, introduced in
\cite{vdhi06}, that maps states $\ket{a}, a\in\F$, to
\[
\sum_{b\in \F}\omega^{\Tr(ab)}\ket{b}\,,
\] 
where $\Tr$ is the trace map from $\F$ to its prime field $\F_p$ and
$\omega=e^{2\pi i/p}$ is a $p$th root of unity. Here, $p$ is the
characteristic of $\F$, that is, $p$ is the prime for which
$|\F|=p^\alpha$ with a positive integer $\alpha$. Then for an element
$a\in\F$ the trace is $\Tr(a)=\sum_{j=0}^{\alpha-1} a^{p^j}$. A
polynomial time approximate implementation of the Fourier transform
of $\F$ is given in \cite{vdhi06}. Here and in the following, we ignore the
error coming from this approximation, because it can be made arbitrarily small
with only a small overhead. This transform maps our
state~(\ref{eq:levelsup2}) to
\[
\sum_{y\in\F}\omega^{\Tr (yw)}
\sum_{x\in\F}\omega^{\Tr(y\nu x +y\mu x^2)}\ket{y}\ket{x}\,.
\]
We measure the first register and obtain the following state (up to a
global phase, which we omit)
\begin{equation}
\label{eq:phases}
\sum_{x\in\F}\omega^{\Tr(y\nu x+y\mu x^2)}\ket{x}
\end{equation}
with uniformly random $y\in\F$. If the term $y\mu x^2$ were missing
from the exponent in the coefficient of $\ket{x}$ in the
state~(\ref{eq:phases}), then $y\nu$ and also $\nu$ could be obtained
by applying the inverse Fourier transform of $\F$.

Motivated by this observation, our goal is to eliminate the quadratic
term from the exponent. To this end, we take three states of the form
(\ref{eq:levelsup2}) with possibly different values $w$ and we apply
the Fourier transform and the measurement independently to them. 
In
principle, we could also consider two states, but taking three
states allows us to apply directly the results of \cite{ISS07} and~\cite{vdW} to simplify the presentation.
In
more detail, we start with the product state
\begin{equation*}
\sum_{
x_1,x_2,x_3 \in \F}\ket{w_1+\nu x_1 +\mu x_1^2}
\ket{w_2+\nu x_2 +\mu x_2^2}\ket{w_3+\nu x_3 +\mu x_3^2}
\ket{x_1,x_2,x_3}
\end{equation*}
and the result is the state
\[
\sum_{x_1,x_2,x_3\in\F}\omega^{\Tr(\nu(y_1x_1+y_2x_2+y_3x_3)
+\mu(y_1x_1^2+y_2x_2^2+y_3x_3^2))}\ket{x_1,x_2,x_3}
\]
for uniformly random $y_1,y_2,y_3\in \F$, which are known
to us as a result of the measurements.
For brevity, we write this state as
\[
\sum_{x_1,x_2,x_3\in\F}\omega^{\Tr(e(x_1,x_2,x_3))}\ket{x_1,x_2,x_3}\,,
\]
where
\[
e(x_1,x_2,x_3)=
\nu(y_1x_1+y_2x_2+y_3x_3)
+\mu(y_1x_1^2+y_2x_2^2+y_3x_3^2)\,.
\]
We abort if any of $y_1,y_2,y_3$ happens to be zero. 
If none of them
is zero, we produce the superposition 
$\sqrt{1/|\F|}\sum_{x\in\F}\ket{x}$ in 
a fourth register. The result is
\[
\sum_{x_1,x_2,x_3,x\in\F}\omega^{\Tr(e(x_1,x_2,x_3))}\ket{x_1,x_2,x_3}\ket{x}\,.
\]
Then, with appropriately chosen elements $\delta_1,\delta_2,\delta_3$
(see below), we subtract $\delta_ix$ from the $i$th register for
$i=1,2,3$. The result is
\[
\sum_{x_1,x_2,x_3,x\in\F}\omega^{\Tr(e(x_1,x_2,x_3))}
\ket{x_1-\delta_1x,x_2-\delta_2x,x_3-\delta_3x}\ket{x}\,,
\]
which is in turn equal to the state
\[
\sum_{x_1,x_2,x_3,x\in\F}\omega^{\Tr(e'(x_1,x_2,x_3))}
\ket{x_1,x_2,x_3}\ket{x}
\]
with
\begin{eqnarray*}
e'(x_1,x_2,x_3) & = & e(x_1+\delta_1 x,x_2 +\delta_2x, x_3 +\delta_3 x) \\
& = &
\nu[y_1(x_1+
\delta_1x)+y_2(x_2+\delta_2x)+y_3(x_3+\delta_3x)]
\\
& &
+\mu[y_1(x_1+\delta_1x)^2+y_2(x_2+\delta_2x)^2+y_3(x_3+\delta_3x)^2] \\
& = &
\nu a(x_1,x_2,x_3)+\mu b(x_1,x_2,x_3) +\nu c x+\mu d(x_1,x_2,x_3)x \\
& &
+ \mu Q x^2,
\end{eqnarray*}
where 
\begin{eqnarray*}
a(x_1,x_2,x_3)&=&y_1x_1+y_2x_2+y_3x_3\\
b(x_1,x_2,x_3)&=&y_1x_1^2+y_2x_2^2+y_3x_3^2\\
c&=&y_1\delta_1+y_2\delta_2+y_3\delta_3\\
d(x_1,x_2,x_3)&=&2(y_1\delta_1x_1+y_2\delta_2x_2+y_3\delta_3x_3)\\
Q&=&y_1\delta_1^2+y_2\delta_2^2+y_3\delta_3^2
\end{eqnarray*}
We choose $\delta_1,\delta_2,\delta_3$ in such a way that the exponent of the
coefficient of $\ket{x_1,x_2,x_3}\ket{x}$ will become linear in $x$
for every $x_1,x_2,x_3$. That is, we want to have $Q=0$. Using the Las
Vegas method of \cite{ISS07} or the deterministic algorithm of
\cite{vdW}, we can find in time ${\rm polylog}( |\F|)$ three elements
$\delta_1,\delta_2,\delta_3$, that are not all zero, such that
$y_1\delta_1^2+y_2\delta_2^2+y_3\delta_3^2=0$. Then the state we have
equals
\[
\sum_{x_1,x_2,x_3,x\in\F}\omega^{
\Tr(\nu a(x_1,x_2,x_3)+\mu b(x_1,x_2,x_3)+
\nu c x + \mu d(x_1,x_2,x_3)x)}\ket{x_1,x_2,x_3}\ket{x}\,.
\]
We measure the first three registers. Then
we obtain the state
\[
\omega^{\Tr(\nu a+\mu b)}\sum_{x\in\F}\omega^{
\Tr(\nu c x + \mu d x)}\ket{x}\,,
\]
where $a=a(x_1,x_2,x_3)$, $b=b(x_1,x_2,x_3)$, and $d=d(x_1,x_2,x_3)$
for uniformly random $x_1,x_2,x_3\in \F$. We abort if $d$ becomes
zero. Note
that $d$ is linear in $x_1,x_2,x_3$ and that none of the $y_i$ 
are zero and that not all of the $\delta_i$ are zero. Hence, $d$ is
only zero with probability $1/|\F|$. 
Observe that here we use the assumption that the characteristic of 
$\F$ is odd, otherwise $d$ would always be zero by definition.
If $d$ is nonzero, we apply
the inverse Fourier transform of $\F$ and obtain the state $\ket{\nu
 c+\mu d}$ up to some phase. After a measurement, we find the value
$g=\nu c+\mu d.$

This way, we obtain a proper linear constraint (as $d\neq 0$) for the
unknown parameters $\mu$ and $\nu$, because the values $c$, $d$ and
$g$ are known to us. The probability that the procedure goes through
is at least $\left(1-1/|\F|\right)^4$. 
The cases when it is aborted
are the cases when one of the values $y_1,y_2$ or $y_3$ is zero or
when $d$ becomes zero.

As $d$ is nonzero, we can substitute $\frac{g}{d}-\frac{c}{d}\nu$ for $\mu$. 
With this
knowledge, our remaining input states are of the form
\begin{equation*}
\sum_{x\in \F}
\ket{w+\nu x + \left( \frac{g}{d} - \frac{c}{d}\nu \right)x^2}\ket{x}
\end{equation*}
By subtracting $\frac{g}{d}x^2$ from the first register,
these become states of the form
\begin{equation}
\label{eq:levelsup3}
\sum_{x\in \F}
\ket{w+\nu x - \frac{c}{d}\nu x^2}\ket{x}
\end{equation}
The Fourier transform of such a state is
\[
\sum_{y\in\F}\omega^{\Tr(yw)}
\sum_{x\in\F}\omega^{\Tr(y\nu x-y\frac{c}{d}\nu x^2)}\ket{y}\ket{x}\,,
\]
which, after measuring the first register, becomes
\begin{equation}
\label{eq:phases2}
\sum_{x\in\F}\omega^{\Tr(y\nu x-y\frac{c}{d}\nu x^2)}\ket{x}.
\end{equation}
With a product of three states of the form (\ref{eq:phases2}) with
nonzero $y$, we repeat the procedure outlined above. It turns out that
we again need to find a nontrivial solution of an equation of the form
$y_1\delta_1^2+y_2\delta_2^2+y_3\delta_3^2$ to get the quadratic term
of the exponent eliminated and to obtain a proper linear constraint
for $\nu$. Having determined $\nu$, we can substitute
$\frac{g}{d}-\frac{c}{d}\nu$ for $\mu$. We used six input states to
determine the hidden polynomial $f(x)=\nu x + \mu x^2$ with high
probability.

This procedure can be extended to higher degrees. We will
give a formal description in the following subsections. Before
that, we briefly outline the method for degree 3,
i.e., when $f(x)=\nu x + \mu x^2 + \kappa x^3$. We assume
that the characteristic of the base field is greater than 3.
In this case, we need four states of the form
\[
\sum_{x\in \F} \omega^{\Tr(yf(x))}\ket{x}
\]
in order to get the cubic term in the exponent of the coefficient
eliminated. We obtain such states from the input states by applying
the Fourier transform and then a measurement on the first register. To
accomplish such an elimination, we have to find a nonzero solution of
an equation of the form
$y_1\delta_1^3+y_2\delta_2^3+y_3\delta_3^3+y_4\delta_4^3=0$. The
result is a similar superposition, with a quadratic exponent. From
twelve input states we first collect three states with quadratic
exponents and from these three states we produce a state with a linear
exponent from which we obtain a linear constraint for the unknown
coefficients. We then perform a similar procedure using the next
twelve input states to obtain a further constraint. Eventually, using
36 input states we will be able to determine all the unknown
coefficients with high probability.

\subsection{Statement of the technical generalization}
In this subsection we formulate a technical generalization of the
\HPFGP{}, which is suitable for recursion. Rather than assuming that
the coefficients of the polynomial $f(x)$ are unknown, we assume that
they linearly depend on some unknown parameters. This generalization
makes it possible to work with polynomials whose coefficients satisfy
some already discovered linear constraints. In order to include
problems related to certain instances of the hidden subgroup problem,
we generalize the problem to {\em tuples of polynomials} at the same
time.

In the general setting we have level sets of a multidimensional (i.e.,
vector-valued) function of the form $F(u,x)=u-f(x)$ with
$u=(u_1,\ldots,u_m)$ and
\begin{equation}\label{eq:2}
f(x)=(f_1(x),\ldots,f_m(x)) \;\; {\rm for} \;\;
f_i(x)=\sum_{s=1}^D a_{is}(v)x^s\,,
\end{equation}
where $a_{is}(v)$ are known homogeneous linear functions in the
unknown $r$-dimensional variable $v=(v_1,\ldots,v_r)$. That is, we
have
\[
a_{is}(v)=\sum_{j=1}^ra_{isj}v_j\,.
\]

\begin{theorem}
\label{thm:ghpgp}
Let $m$, $r$ and $D$ be constants and let $f:\F\rightarrow \F^m$ be a
function as defined in (\ref{eq:2}). Then there is a quantum algorithm
which, given $O(1)$ states of the form
\[
\sum_{x\in \F}\ket{w+f(x)}\ket{x} 
\]
for various unknown vectors $w \in \F^m$, determines the unknown
parameters $v_1,\ldots,v_r$ efficiently.
\end{theorem}

Theorem~\ref{thm:hpgp} is the special case of Theorem~\ref{thm:ghpgp}
with $m=1$, $r=D$, and $a_{1s}(v)=v_{s}=\nu_s$. As a direct
consequence, we also obtain the following result regarding a
multidimensional generalization of the \HPFGP{} to which the
hidden subgroup problem in $\Z_p^m\rtimes \Z_p$
can be efficiently reduced (see~\cite{DISW11}).

\begin{corollary}
\label{cor:mhpgp}
Let $f(x)=(\sum_{s=1}^D\nu_{1s}x^s,\ldots,\sum_{s=1}^D\nu_{ms}x^s)$.
Then there is a quantum algorithm which, given $O(1)$ states of the
form
\[
\sum_{x \in \F}\ket{w+f(x)}\ket{x}
\]
for various unknown $w\in \F^m$, determines efficiently the unknown
coefficients $\nu_{is}$ for $i=1,\ldots,m$ and $s=1,\ldots,D$.
\end{corollary}

Corollary~\ref{cor:mhpgp} is indeed a special case of
Theorem~\ref{thm:ghpgp} with $r=mD$ and $a_{is}(v)=v_{(i-1)D+s} =
\nu_{is}$. 

\subsection{A high-level description of the algorithm}

The algorithm for Theorem~\ref{thm:hpgp} 
is organized as a recursion on the number $r$ of 
the unknown parameters $v_1,\ldots,v_r$. The recursion 
(described in Subsection~\ref{ss-outer}) is based on 
eliminating one of the parameters by finding a linear equation for them.

The 
procedure for finding a linear equation for the
parameter 
starts with producing many states 
of the form
$$\sum_{x\in\F}\omega^{\Tr(\sum_{j=1}^r
\sum_{s=1}^Dv_jY_{js}x^s)}\ket{x},$$
where $Y_{js}$ ($j=1,\ldots,r$, $s=1,\ldots,D$) are 
elements from $\F$ depending on certain measurements
(see Subsection~\ref{ss-init} for details). During the algorithm
we will work with states of the form above, with less and less
nonzero coefficients $Y_{js}$. To shorten the discussion,
in this subsection we refer to the polynomial 
$\sum_{j=1}^r\sum_{s=1}^Dv_jY_{js}x^s$
as the {\em phase} of the state.

Assume for simplicity that the characteristic $p$ of our field
$F$ is larger than the degree $D$. Then
we make small groups of such states. {From} each group,
using a method similar to what is outlined in Subsection~\ref{ss-smalldeg},
we fabricate a single state in the phase of which one of the 
highest degree terms (i.e., $Y_{jD}v_{j}x^{D}$ for some $j$) gets eliminated,
that is, in the new state the coefficient $Y_{jD}$ becomes
zero. {From} the new states we again form small groups to
make states where further high degree terms get eliminated.
We proceed this way until we get a state where the 
phase has linear terms only, that is, a state of the form
$$\sum_{x\in\F}\omega^{\Tr(\sum_{j=1}^rv_{j}Y_{j1}x)}\ket{x}.$$
Application of the inverse Fourier transform of $\F$ and a measurement 
gives then the value of $\sum_{j=1}^rv_{j}Y_{j1}$,
which can be used as a linear equation for the parameters $v_1,\ldots,v_r$.

It turns out that over a field of characteristic $p$ smaller than
$D$ the terms of degree $s$, where $s$ is a power of $p$,
cannot be eliminated from the phase using a method
similar to that of Subsection~\ref{ss-smalldeg}. 
Fortunately, such a method is still applicable to produce
a state in the phase of which all the terms whose
degree is not a power of $p$ are eliminated, see 
the first part of Subsection~\ref{sec:eliminate}. 
The remaining high degree terms are eliminated by using a 
slightly different technique
based on groups of size 2, see the second part of
Subsection~\ref{sec:eliminate} for
details.

\subsection{The outer loop}
\label{ss-outer}
The algorithm for Theorem~\ref{thm:ghpgp} uses a recursion by $r$.
The main ingredient of the recursion is a procedure (described in the
following two subsections), which, using sufficiently many input
states, finds with high probability a linear equation
\begin{equation}\label{eq:3}
\sum_{j=1}^r\alpha_jv_j=\beta
\end{equation}
that is satisfied by the unknown parameters $v_j$, where
$\alpha_1,\ldots,\alpha_r,\beta\in \F$ and at least one $\alpha_i$ is
nonzero.

Assume without loss of generality that we obtained such an equation with $\alpha_{r}\neq
0$. Then we substitute $v_r= \frac{\beta}{\alpha_r}-
\sum_{j=1}^{r-1}\frac{\alpha_j}{\alpha_r}v_j$. We have
\[
f_i(x)= \sum_{s=1}^D\sum_{j=1}^{r} a_{isj}v_jx^s= \sum_{s=1}^D
\sum_{j=1}^{r-1} (a_{isj}-\frac{\alpha_j}{\alpha_r}a_{isr})v_jx^s
+\frac{\beta}{\alpha_r}\sum_{s=1}^D a_{isr}x^s\,.
\]
We apply the recursion to the hidden function
$f^*=(f_1^*,\ldots,f_m^*)$ with
\[
f_i^*(x)=f_i(x)- \frac{\beta}{\alpha_r}\sum_{s=1}^Da_{isr}x^s\,.
\]
Note that the coefficients of $x^s$ only depend on the unknown
$v_1,\ldots,v_{r-1}$ and that the level sets of $u-f^*(x)$ and
$u-f(x)$ differ only by a shift with
\[
\left(\frac{\beta}{\alpha_r}\sum_{s=1}^Da_{1sr}x^s,\ldots,
\frac{\beta}{\alpha_r}\sum_{s=1}^Da_{msr}x^s \right)\,.
\]
This means that if $\ket{u_1}\cdots\ket{u_m}\ket{x}$ belongs to the
level set of a certain value $w$ of the function $u-f(x)$ then we
subtract $\frac{\beta}{\alpha_r}\sum_{s=1}^Da_{isr}x^s$ from the $i$th
register for all $i=1,\ldots,m$ and this leads to an element
$\ket{u_1^*}\cdots\ket{u_m^*}\ket{x}$ of the level set of the function
$u-f^*(x)$ corresponding to $w$. We determine the values
$v_1,\ldots,v_{r-1}$ by recursion from which $v_r$ can be computed
using our linear constraint. In the base case of the recursion the
main procedure gives us a linear constraint for the only unknown
$v_1$, which allows us to determine its value easily.

\subsection{The initial stage of the inner procedure}
\label{ss-init}
The level set superpositions for $F(u,x)=u-f(x)$ are states of the
form
\begin{equation*}
\sum_{x\in \F}
\ket{w+f(x)}\ket{x}=
\sum_{x\in \F}\ket{w_1+\sum_{s=1}^D\sum_{j=1}^ra_{1sj}v_{j} x^s}
\cdots
\ket{w_m+\sum_{s=1}^D\sum_{j=1}^ra_{msj}v_{j}x^s}\ket{x}
\end{equation*}
for various vectors $w=(w_1,\ldots,w_m)\in \F^m$.
We apply the quantum Fourier transform of the field $\F$ independently
on all of the first $m$ registers and obtain the state
\[
\sum_{y\in \F^m} \sum_{x\in \F} \omega^{\Tr(\sum_{i=1}^m y_iw_i+
\sum_{i=1}^m y_i\sum_{j=1}^rv_{j}\sum_{s=1}^Da_{isj}x^s)}\ket{y}\ket{x}\,.
\]
Then we measure $y$ and obtain the state
\[
\omega^{\Tr(\sum_{i=1}^my_iw_i)}
\sum_{x\in \F}
\omega^{\Tr(\sum_{i=1}^my_i\sum_{j=1}^rv_{j}\sum_{s=1}^Da_{isj}x^s)}\ket{x}
\]
with uniformly random $y=(y_1,\ldots,y_m)\in\F^m$. After forgetting
the global phase, we have
\begin{equation}\label{eq:4}
\sum_{x\in\F}\omega^{\Tr (\sum_{i=1}^my_i\sum_{j=1}^rv_{j}\sum_{s=1}^Da_{isj}x^s)}\ket{x}
=
\sum_{x\in\F}\omega^{\Tr(\sum_{j=1}^rv_{j}
\sum_{s=1}^DY_{js}x^s)}\ket{x}\,,
\end{equation}
where
\[
Y_{js}=\sum_{i=1}^m y_ia_{isj}\;\;{\rm for} \;\; s=1,\ldots,D\;\; {\rm
 and}\;\; j=1,\ldots,r\,.
\]
We keep this state only if not all the coefficients $Y_{js}$ are zero.
Provided not all the parameters $a_{isj}$ are zero, this happens with a
probability of at least $\frac{|\F|-1}{|\F|}$. In the following, we
use several states of type (\ref{eq:4}) to obtain similar states, but
where the highest-degree term $Y_{js}x^s$ of the phase gets
eliminated. We will accomplish this elimination with an iterative
method, which is described in the next subsection. Eventually, we
obtain a state with linear terms only. Such a state will be used to
set up a linear equation for the unknown parameters $v_j$.

\subsection{Eliminating high degree terms from the phase}\label{sec:eliminate}
Here we show how to eliminate the high degree terms from the
phase. We consider terms whose degree is a power of $p$ and
terms whose degree is not a power of $p$ separately, because the
characteristic affects the solvability of equations.

First we describe an iterative procedure which eliminates the
terms whose degree
is not a power of the characteristic $p$ of $\F$. The iteration is
controlled by a single tuple $(n_1,\ldots,n_r)$ of integers between $1$ and
$D$ and we initialize it with
\[
(n_1,\ldots, n_r)=(D,\ldots, D)\,.
\]
A step of the iteration receives $\ell\leq D+1$ states of the form
\[
\sum_{x\in\F}\omega^{\Tr(\sum_{j=1}^r v_j
 \sum_{s=1}^{D}Y_{js}x^s)}\ket{x}\,,
\]
where we have
\[
Y_{js}=0\;\mbox{whenever $s>n_j$ and $s$ is not a power of $p$}\,.
\]
In the case that not all $n_j$ are equal to 1, we define $j_0$ to be
the smallest index $j$ such that $n_j>1$ and the procedure fabricates
a state of the form
\[
\sum_{x\in\F}\omega^{\Tr(\sum_{j=1}^r v_j
 \sum_{s=1}^{D}Y^*_{js}x^s)}\ket{x}\,,
\]
where not all $Y_{js}^*$ are equal to zero but
\[
Y^*_{js}=0\;\mbox{whenever $s>n_j$ and $s$ is not a power of $p$,}
\]
and additionally
\[
Y^*_{j_0,n_{j_0}}=0\;\mbox{if $n_{j_0}$ is not a power of $p$}\,.
\]
The step is trivial if $n_{j_0}$ is a power of $p$, or 
if one of the $\ell$ states, say the ${j}^{\textup{th}}$ state, 
already satisfies that $n_j>1$ and $Y_{j,n_{j}}=0$.
In the
following, we describe the details of the step for the remaining
case.

The input to the iterative step consists of the elements
$Y_{js}^{(i)}\in\F$ for $i=1,\ldots,\ell$ and $j=1,\ldots,r$ and
$s=1,\ldots,D$. We also have the product state
\begin{equation}\label{eq:5}
\sum_{(x_1,\ldots,x_\ell)\in\F^\ell} \omega^{\Tr(\sum_{j=1}^r v_j
 \sum_{s=1}^{D}\sum_{i=1}^\ell
 Y_{js}^{(i)}x_i^s)}\ket{x_1,\ldots,x_\ell}\,,
\end{equation}
where we have
\[
Y^{(i)}_{js}=0\;\mbox{whenever $s>n_j$ and $s$ is not a power of $p$}\,,
\]
but 
\[
Y^{(i)}_{j_0,n_{j_0}}\neq 0 \mbox{~for $i=1,\ldots,\ell$}\,.
\]
We assume that $n_{j_0}=p^\beta b$, where $b$ is an integer that
is coprime to $p$ and greater than 1.

We start with appending $\sum_{x\in \F}\ket{x}$
to the product state (\ref{eq:5}).
This way we obtain
\[
\sum_{(x_1,\ldots,x_\ell)\in\F^\ell}\sum_{x\in\F}
\omega^{\Tr(\sum_{j=1}^r v_j \sum_{s=1}^{D}\sum_{i=1}^\ell
Y_{js}^{(i)}x_i^s}\ket{x_1,\ldots,x_\ell}\ket{x}\,.
\]
Next we choose elements $\delta_1,\ldots,\delta_\ell\in\F$, which are not all
equal to zero, such that
\[
\sum_{i=1}^\ell \left(Y^{(i)}_{j_0,n_{j_0}}\right)^{p^{-\beta}}\delta_i^{b}=0\,.
\]
Using $\ell>n_{j_0}=p^\beta b\geq b$, this can be done in 
deterministic polynomial time by~\cite{vdW}. For later use we ensure that
the tuple $(\delta_1,\ldots,\delta_\ell)$ depends
only on the ratios between the parameters $Y_{js}^{(i)}$. This
can be done by normalizing the input elements
$(Y_{j_0n_{j_0}}^{(i)})^{p^{-\beta}}$ for \cite{vdW} such that the first nonzero coefficient
becomes one. 

We subtract $\delta_ix$ from the $i$th register
and substitute $x_i$ for $x_i-\delta_ix$ to obtain
the state
\begin{equation}\label{eq:prior6}
\sum_{(x_1,\ldots,x_\ell)\in\F^\ell}\sum_{x\in \F}
\omega^{\Tr(\sum_{j=1}^r v_j \sum_{s=1}^{D}\sum_{i=1}^\ell
Y_{js}^{(i)}(x_i+\delta_ix)^s)}\ket{x_1,\ldots,x_\ell}\ket{x}\,.
\end{equation}
We measure $x_1,\ldots,x_\ell$ and forget the global phase. Hence, we obtain
the state
\begin{equation}\label{eq6}
\sum_{x\in \F}
\omega^{\Tr(\sum_{j=1}^r v_j \sum_{k=1}^{D}
Y_{jk}^*x^k)}\ket{x}\,,
\end{equation}
where
\[
Y_{jk}^*=\sum_{s=k}^{D}\binom{s}{k}\sum_{i=1}^\ell
Y_{js}^{(i)}x_i^{s-k}\delta_i^k\,.
\]
Since (\ref{eq:prior6}) is a uniform superposition over all choices of
$x$ and $x_1,\ldots,x_\ell$, except with different phases, the
measurement produces uniformly random $x_1,\ldots,x_\ell$.

Note that if $s$ is a power of the characteristic $p$
and if $s>k$
then the integer $\binom{s}{k}$
is divisible by $p$. Therefore, the terms
\[
\binom{s}{k}\sum_{i=1}^\ell
Y_{js}^{(i)}x_i^{s-k}\delta_i^k
\]
are zero. For $s>n_j$, which is not a power of $p$, the terms
\[
\binom{s}{k}\sum_{i=1}^\ell Y_{js}^{(i)}x_i^{s-k}\delta_i^k
\]
are zero as well, because the parameters $Y_{js}^{(i)}$ are all zero. 
This shows that we have
\[
Y^*_{jk}=0\;\mbox{whenever $k>n_j$ and $k$ is not a power of $p$}\,.
\]
We also have
\[
Y^*_{j_0,n_{j_0}}=\sum_{i=1}^\ell Y^{(i)}_{j_0,n_{j_0}}\delta_i^{n_{j_0}}=
\left(\sum_{i=1}^\ell
\left(Y^{(i)}_{j_0,n_{j_0}}\right)^{p^{-\beta}}\delta_i^{b}\right)^{p^\beta}=0
\]
by the choice of the $\delta_j$.
Furthermore, the equation
\[
Y_{j_0,p^\beta(b-1)}^*=
\sum_{s=p^\beta(b-1)}^{p^\beta b}\binom{s}{p^\beta(b-1)}\sum_{i=1}^\ell
Y_{j_0,s}^{(i)}x_i^{s-p^\beta(b-1)}\delta_i^{p^\beta(b-1)}
\]
shows that 
$Y_{j_0,p^\beta(b-1)}^*$
is a polynomial of degree $p^\beta$ in the
variables $x_1,\ldots,x_\ell$. The 
homogeneous part of degree $p^\beta$ is
\[
\binom{n_{j_0}}{p^\beta (b-1)}
\sum_{i=1}^\ell
Y_{j_0,n_{j_0}}^{(i)}x_i^{p^\beta}\delta_i^{p^\beta(b-1)}\,.
\]
As none of the parameters $Y_{j_0,n_{j_0}}^{(i)}$ is zero, the
coefficients $Y_{j_0,n_{j_0}}^{(i)}\delta_i^{p^\beta(b-1)}$ are not
all zero. From this we conclude,
using the fact that the binomial coefficient $$\binom{n_{j_0}}{p^\beta (b-1)}= \binom{p^\beta
b}{p^\beta (b-1)}=\binom{p^\beta b}{p^\beta}$$ is not divisible by
$p$ as $1<b<p$,
that $Y_{j_0,p^\beta(b-1)}^*$ is not
identically zero when considered as a polynomial in the variables
$x_1,\ldots,x_\ell$. As the measurements give us uniformly random
values $x_1,\ldots,x_\ell$, we know by the Schwartz--Zippel lemma that
$Y_{j_0,p^\beta(b-1)}^*$ will be nonzero with a probability of at
least
\[
\frac{|\F|-p^\beta}{|\F|}\geq \frac{|\F|-D}{|\F|}\,.
\]
This shows that not all the new coefficients $Y_{js}^*$ for $s \leq
n_j$ will be zero with a probability of at least
$\frac{|\F|-D}{|\F|}$, because $Y_{j_0,p^\beta(b-1)}^*$ is one of
these coefficients.  With high likelyhood, we have produced a state of
the form (\ref{eq6}) for a set of new coefficients $Y_{js}^*$ that are
known linear polynomials in the original coefficients~$Y_{js}^{(i)}$.

After eliminating the terms whose degrees are not a power of $p$, we
now explain how to deal, in the remaining rounds
of the iteration, with terms whose degrees are a power of $p$.  Our intention is the following. From several states
of the form 
\[
\sum_{x\in\F}\omega^{\Tr(\sum_{j=1}^rv_j\sum_{t=0}^{d}
Z_{jt}x^{p^t})}\ket{x}\,,
\]
where $d={\lfloor \log_p D\rfloor}$
and not all of the coefficients $Z_{jt}$ are equal to zero,
we produce a state that has only linear terms.
Now the iteration is controlled by a single tuple
$(n_1,\ldots,n_r)$ of integers between $0$ and $d$,
which are not all equal to zero, and the iteration
starts with the tuple
\[
(n_1,\ldots, n_r)=(d,\ldots,d)\,.
\]
A step of the iteration receives the coefficients
$Z^{(i)}_{jt}$ for $j=1,\ldots, r$ and $t=0,\ldots,d$ 
and $i=1,2$ along with the two states 
\[
\sum_{x_i\in\F}\omega^{\Tr(\sum_{j=1}^rv_j\sum_{t=0}^{d}
Z^{(i)}_{jt}x_i^{p^t})}\ket{x_i}
\]
such that for both $i=1,2$ the coefficients $Z^{(i)}_{jt}$ are not all
zero but $Z^{(i)}_{jt}=0$ whenever $t>n_j$. Let $j_0$ be the smallest
integer $j$ such that $n_j>0$. We simply pass the appropriate state to
the next round if either $Z^{(1)}_{j_0,n_{j_0}}$ or $Z^{(2)}_{j_0,n_{j_0}}$ is
zero.

Otherwise let us assume first that there exist two pairs $(t_1,j_1)$
and $(t_2,j_2)$ with $t_1\neq t_2$ such that $Z^{(1)}_{j_1,t_1}\neq 0$ and
$Z^{(2)}_{j_2,t_2}\neq 0$. Then we abort if there is an element $z\in\F$
such that $Z^{(2)}_{jt}=z^{p^t}Z^{(1)}_{jt}$ for all $j=1,\ldots,r$ and for
all $t=0,\ldots,d$. Otherwise we append $\sum_{x\in \F}\ket{x}$ to the
product state
\[
\sum_{(x_1,x_2)\in\F^2}
\omega^{\Tr(\sum_{j=1}^r v_j \sum_{t=0}^{d}
(Z_{jt}^{(1)}x_1^{p^t}+Z_{jt}^{(2)}x_2^{p^t}))}\ket{x_1,x_2}
\]
and we obtain
\[
\sum_{x\in\F}\sum_{(x_1,x_2)\in\F^2}
\omega^{\Tr(\sum_{j=1}^r v_j \sum_{t=0}^{d}
(Z_{jt}^{(1)}x_1^{p^t}+Z_{jt}^{(2)}x_2^{p^t}))}\ket{x_1,x_2}\ket{x}\,.
\]
Then we set
\[
\delta_1=1 \;\; {\rm and} \;\;
\delta_2=-\left(\frac{Z^{(1)}_{j_0,n_{j_0}}}{Z^{(2)}_{j_0,n_{j_0}}}\right)^{p^{-n_{j_0}}}
\]
and we
subtract
$\delta_ix$ from the $i$th register. After 
substituting $x_i$ for $x_i-\delta_ix$ we obtain 
the state
\[
\sum_{x \in \F} \sum_{(x_1,x_2)\in\F^2}
\omega^{\Tr(\sum_{j=1}^r v_j \sum_{t=0}^{d}
(Z_{jt}^{(1)}(x_1+\delta_1x)^{p^t}+Z_{jt}^{(2)}
(x_2+\delta_2x)^{p^t}))}\ket{x_1,x_2}\ket{x}\,.
\]
We measure $x_1,x_2$ and after forgetting the global phase we obtain
the state
\[
\sum_{x\in \F}
\omega^{\Tr(\sum_{j=1}^r v_j \sum_{t=0}^{d}
Z_{jt}^*x^{p^t})}\ket{x}\,,
\]
where we have 
\[
Z_{jt}^*=Z^{(1)}_{jt}\delta_1^{p^t}+Z^{(2)}_{jt}\delta_2^{p^t}\,.
\]
By the choice of $\delta_1$ and $\delta_2$,
we have $Z_{j_0,n_{j_0}}^*=0$, and by the 
assumption 
\[
Z_{jt}^{(2)} \not = - Z_{jt}^{(1)} \left(\frac{\delta_1}{\delta_2} \right)^{p^t}
\]
for some $j$ and $t$ not all 
$Z_{jt}^*$ are equal to zero. Again,
$\delta_1$ and $\delta_2$ depend only on
the ratios between the parameters $Z^{(i)}_{jt}$.

If all $Z^{(1)}_{jt}$ are zero except for $t=n_{j_0}$
then we replace $x_1$ with $x_1^{p^{-n_{j_0}}}$
and finish the iteration with the state
\[
\sum_{x_1\in\F}\omega^{\Tr(\sum_{j=1}^r v_j Z^{(1)}_{j,n_{j_0}}x_1)}\ket{x_1}
\]
and the inverse Fourier transform gives us the
sum 
\[
\sum_{j=1}^r v_j Z_{j,n_{j_0}}^{(1)}
\]
for the unknown $v_1, \ldots, v_r$. 

The iterative procedure above, starting with $L=O(1)$ states of the form
(\ref{eq:4})
with uniformly random $y_1,\ldots,y_m$, 
constructs a state of the form
\begin{equation}\label{eq:7}
\sum_{x\in\F}\omega^{\Tr(\sum_{j=1}^r\alpha_jv_jx)}\ket{x}\,,
\end{equation}
where not all of the $\alpha_j$ are zero, with high success
probability in time ${\rm polylog} (|\F|)$. We apply the inverse of
the quantum Fourier transform of $\F$ to the state (\ref{eq:7}) and
obtain the state
\[
\ket{\sum_{j=1}^n\alpha_jv_j}\,.
\]
When we measure this state and denote the result by $\beta$, then we
have a linear constraint of the form (\ref{eq:3}) for the unknown
$v_j$.

The probability of abortion, i.e., there is a $z \in \F$ with
$Z^{(2)}_{jt} = z^{p^t} Z^{(1)}_{jt}$ for all $j$ and $t$, can be
estimated as follows.  First, assume that for a run of the iteration
to compute the state with coefficients $Z_{jt}^{(2)}$ we have tuples
$(y_1^{(i)}, \ldots, y_m^{(i)})$ with $i=1,\ldots, L$ as measurement
results in the beginning. Then for a $\gamma \in \F\setminus\{0\}$ the
iteration for the measurement results $(\gamma y_1^{(i)}, \ldots,
\gamma y_m^{(i)})$ takes the same course and we obtain the
coefficients $\gamma Z_{jt}^{(2)}$, because all $Y_{js}^{(i)}$ and
$Z_{jt}^{(i)}$ are just homogeneous linear combinations.  When we have
$Z_{jt}^{(2)} = z^{p^t} Z_{jt}^{(1)}$ for all $j$ and $t$, then for
$\gamma \not = 0$ we would have $\gamma Z_{jt}^{(2)} = z^{p^t}
Z_{jt}^{(1)}$ and this cannot hold for all $t$ when $\gamma \not = 1$.
Therefore, when we consider a fixed first state $Z^{(1)}_{jt}$, then
for each measurement result for the second collection of states that
leads to an abortion because of $Z^{(2)}_{jt} = z^{p^t} Z^{(1)}_{jt}$,
there are at least $|\F|-2$ possibilities of other measurement results
that do not lead to abortion.  The normalization in both of the iteration
steps ensures that for every $w\in \F$ the probability of obtaining a
state with $\gamma Z^{(2)}_{jt}$ instead of $Z^{(2)}_{jt}$ for every
$t$ and $j$ are the same for every $\gamma \in
\F\setminus\{0\}$. 
Therefore, if $Z^{(1)}_{j_1,t_1}$ and $Z^{(2)}_{j_2,t_2}$ are nonzero
for two pairs $(j_1,t_1)$ and $(j_2,t_2)$, the conditional property of
having $Z^{(2)}_{jt}=z^{p^t}Z^{(1)}_{jt}$ is at most
$\frac{1}{|\F|-1}$.

This finishes the description of the algorithm for
Theorem~\ref{thm:ghpgp}.

\section{Application: Special hidden polynomials}\label{sec:hpps}

In this section we prove Theorem~\ref{thm:hpps}.
To simplify the notation, we define the cardinality 
of the level set of the function $F(x,y)=g(y)-f(x)$ 
corresponding to $w \in \F$ to be 
\[
M_F(w)=\#\{(x,y)\in\F^2:g(y)-f(x)=w\}
\]
and in a similar manner 
\[
m_g(w)=\#\{y\in\F:g(y)=w\}
\]
to be the cardinality of the level set of $g(y)$ corresponding
to $w \in \F$. We trivially have that
\[
\sum_{w\in\F}M_F(w)=|\F|^2
\]
and that
\[
\sum_{x\in\F}m_g(f(x)+w)=M_F(w)
\]
for every $w\in\F$.
The quantum procedure starts with the uniform superposition
\[
\frac{1}{|\F|}
\sum_{x,y\in\F}\ket{y}\ket{x}\ket{0}
\]
and we apply the oracle to produce the state
\[
\frac{1}{|\F|}
\sum_{x,y\in\F}\ket{y}\ket{x}\ket{{\cal E}(F(x,y))}\,.
\]
Then we measure the third register and we obtain the result
${\cal E}(w)$ for a $w \in \F$ with probability $M_F(w)/|\F|^2$.
The resulting state of the first two registers is 
\[
\ket{\Phi_w}=
\frac{1}{\sqrt{M_F(w)}}
\sum_{ \{ (x,y) \in \F^2: g(y)-f(x)=w\}} \ket{y}\ket{x}\,.
\]

Note that this state is similar to the level set superposition of the
function $y-f(x)$ corresponding to the value $w$. To exploit this
connection, we make use of the unitary map ${\cal U}_g$ which
maps
$\ket{z}$ to $\frac{1}{\sqrt{m_g(z)}}\sum_{y:g(y)=z}\ket{y}$ for $z \in \F$. 
The case that $g(y)=z$ has no solution cannot occur in our algorithm
and therefore we set the result of ${\cal U}_g$ to a special state in this case.
We can implement ${\cal U}_g$ as follows.
\begin{enumerate}
\item Compute
$S_z=\{y:g(y)=z\}$ in an ancilla using Berlekamp's root
finding algorithm~\cite{Berlekamp68}.
\item Produce the uniform superposition
$\ket{S_z}=\frac{1}{\sqrt{m_g(z)}}\sum_{y \in S_z}\ket{y}$ in another ancilla.
\item Erase the first ancilla by undoing the first step.
\item Swap $\ket{z}$ with $\ket{S_z}$.
\item Erase the ancilla holding $\ket{z}$
by evaluating $g$ on $\ket{S_z}$.
\end{enumerate}

We apply ${\cal U}_g$ to the first register of $\ket{\Phi_w}$ and
obtain the state
\[
\ket{\Psi_w}=\frac{1}{\sqrt{M_F(w)}}
\sum_{x\in\F}
\sqrt{m_g(f(x)+w)}\ket{w+f(x)}\ket{x}\,.
\]
Let $\ket{\Lambda_w}$ stand for the state corresponding to the level set
$\{x\in\F:u-f(x)=w\}$ of the function $u-f(x)$, that is,
\[
\ket{\Lambda_w}=\frac{1}{\sqrt{|\F|}}\sum_{x\in\F}\ket{w+f(x)}\ket{x}\,.
\]
Then the scalar product of $\ket{\Psi_w}$ with $\ket{\Lambda_w}$ is
\[
\langle\Psi_w | \Lambda_w \rangle=
\frac{1}{\sqrt{M_F(w)|\F|}}\sum_{x\in \F}\sqrt{m_g(f(x)+w)}\,.
\]
We apply the inequality $m_g(f(x)+w)\leq \deg (g)=D'$ to obtain
\[
\langle\Psi_w|\Lambda_w\rangle\geq
\frac{1}{\sqrt{M_F(w)|\F|}}\sum_{x\in
 \F}\frac{1}{\sqrt{D'}}{m_g(f(x)+w)} = \frac{\sqrt{M_F(w)}}{\sqrt
 {D'|\F|}}\,.
\]
The expected value of $M_F(w)$ is
\[
\sum_{w\in\F}\frac{M_F(w)}{|\F|^2}M_F(w)\geq
\frac{1}{|\F|}\left(\frac{1}{|\F|}\sum_{w\in\F}M_F(w)\right)^2
=\frac{1}{|\F|}\left(\frac{1}{|\F|}|\F|^2\right)^2=|\F|.
\] 
As the maximum possible value for $M_F(w)$ is $D'|\F|$ and as $D'$ is
a constant, we have that $M_F(w)$ is at least $\frac{1}{2}|\F|$ with a
probability that is lower bounded by a positive constant. This implies
that the scalar product of $\ket{\Psi_w}$ with $\ket{\Lambda_w}$ is
also at least another positive constant. Therefore, if we apply the
algorithm of Theorem~\ref{thm:hpgp} to $O(1)$ states of the form
$\ket{\Phi_w}$ instead of $\ket{\Lambda_w}$, we still have a
probability, which is lower bounded by a positive constant, that the
algorithm determines the coefficients of $f(x)$ correctly.

Using the above procedure we obtain a guess $f_0(x)$ for $f(x)$. We
can test correctness of such a guess using the oracle as follows. As
$\#\{(x,y)\in\F^2:g(y)-f_0(x)=w\}$ is $|\F|$ on average and as the
maximum is $D'|\F|$, for a uniformly random $w$ this number will be at
least $|\F|/2$ with a probability of at least $\frac{1}{2D'}$. For
such a $w$, at least for $\frac{|\F|}{2D}$ values $x$, the equation
$g(y)=f_0(x)+w$ has at least one solution $y$ which can be obtained by
Berlekamp's root finding algorithm. Using this strategy, we find in
${\rm polylog}(|\F|)$ time with high probability an element $w\in \F$
and $D+1$ pairs $(x_1,y_1),\ldots,(x_{D+1},y_{D+1})$ from $\F^2$ such
that $x_i\neq x_j$ holds whenever $i\neq j$ and $f_0(x_i)-g(y_i)=w$
for $i=1,\ldots,D+1$. We call the oracle for the pairs $(x_i,y_i)$ and
check if it returns the same value for all $i=1,\ldots,D+1$. If it
does not then it is impossible that $f$ and $f_0$ are the same up to
constant term. However, if it does then we have $f(x_i)=g(y_i)+w'$ for
$i=1,\ldots,D+1$ and for some $w'\in \F$, whence
$f(x_i)=f_0(x_i)+w'-w$ for $i=1,\ldots,D+1$. From this,
$f(x)=f_0(x)+w'-w$ follows for every $x\in \F$ because $f$ has degree
at most $D$. This completes the proof of Theorem~\ref{thm:hpps}.

\section*{Acknowledgments}
\noindent
Most of
this work was conducted at the Centre
for Quantum Technologies (CQT) in Singapore, and partially funded by
the Singapore Ministry of Education and the National Research
Foundation, also through the Tier 3 Grant ``Random numbers from quantum processes".
Research partially supported by the 
Natural Sciences and Engineering Research Council of Canada,
the European Commission IST STREP projects Quantum Computer Science (QCS) 255961
and Quantum Algorithms (QALGO) 600700,
by the French ANR Blanc program under contract ANR-12-BS02-005 (RDAM project), 
and by the Hungarian Scientific Research Fund (OTKA), Grants NK105645 and
K77476.

\end{document}